# When Life gives you LLMs, make LLM-ADE: Large Language Models with Adaptive Data Engineering

Stephen Choi, William Gazeley
IRAI Labs
`{stepchoi, william}@irai.co`

## Abstract

This paper presents the LLM-ADE framework, a novel methodology for continued pre-training of large language models (LLMs) that addresses the challenges of catastrophic forgetting and double descent. LLM-ADE employs dynamic architectural adjustments, including selective block freezing and expansion, tailored to specific datasets. This strategy enhances model adaptability to new data while preserving previously acquired knowledge. We demonstrate LLM-ADE's effectiveness on the TinyLlama model across various general knowledge benchmarks, showing significant performance improvements without the drawbacks of traditional continuous training methods. This approach promises a more versatile and robust way to keep LLMs current and efficient in real-world applications.

## 1 Introduction

Large Language Models (LLMs) have become pivotal in artificial intelligence, celebrated for their capacity to assimilate and utilize extensive general knowledge (Brown et al., 2020; Kojima et al., 2022). These models are trained on broad datasets, enabling them to generate text across various subjects, often demonstrating a broad yet sometimes superficial understanding of information (Anil et al., 2023). However, this breadth can come at the cost of depth, as LLMs may generate inaccurate, "hallucinated" content, particularly in task-oriented dialogues (Bang et al., 2023), and struggle in specialized domains demanding precise knowledge (Shen et al., 2023).

The training of LLMs is resource and time-intensive (Kaplan et al., 2020) and is bounded by a knowledge cut-off date (Cheng et al., 2024), limiting their ability to incorporate up-to-date information. Additionally, these models require extensive data preprocessing and curation, which may be impractical for real-world applications where data are often unprocessed, duplicated, and frequently updated. To address these challenges, a novel training methodology that allows for rapid adaptation to new data without the drawbacks of traditional continuous training methods is needed. This paper introduces the LLM-ADE framework, an innovative approach for the continued pre-training of LLMs that enhances their learning efficiency from specific datasets while preventing issues such as catastrophic forgetting and double descent.

Existing methods like Retrieval Augmented Generation (RAG) provide LLMs with non-parametric memory, which helps but falls short in complex reasoning tasks (Lewis et al., 2020; BehnamGhader et al., 2023). Fine-tuning improves domain adaptation but often yields suboptimal results and struggles with extended contexts (Chung et al., 2022; Anil et al., 2022; Zhou et al., 2023).

Direct improvements to the core structure of LLMs can yield benefits across various applications, reducing the need for extensive downstream task-specific tuning. However, continuous domain specific training, risks diminishing the model's broad applicability and is vulnerable to double descent and catastrophic forgetting, where model performance degrades or essential knowledge is lost (Belkin et al., 2019; Lopez-Paz and Ranzato, 2017). Notably, data duplication during training exacerbates performance issues (Hernandez et al., 2022), and ongoing training can erode general knowledge (Luo et al., 2023a), particularly models in the 1-7 billion parameter range (Luo et al., 2023b).

The LLM-ADE framework (Large Language Models with Adaptive Data Engineering) is designed to meet three critical real-world criteria: 1) process and generate language on any specified dataset, including those of lower quality or with pre-training data overlap, 2) retain general-purpose applicability without catastrophic forgetting, 3) achieve high efficiency in resource utilization and training time. LLM-ADE incorporates a dynamic architectural adjustment strategy, utilizing layer importance techniques to freeze and expand on certain layers, tailored by the specified dataset (corpus) to preserve general knowledge but accommodate new information. The framework's adaptability not only enables it to maintain performance across various domains but also paves the way for a more efficient continuous learning paradigm in LLMs. As such, LLM-ADE provides a promising direction for future research and applications, aiming to make the continuous pre-training of LLMs more accessible, versatile, and robust.

## 2 Related Work

In the field of continuous pre-training (CPT), several approaches have been developed to update language models with new information while mitigating the risk of catastrophic forgetting. Jang et al. (2022) introduced a method for continual knowledge learning in LLMs that focuses on temporal knowledge updates with reduced forgetting. Ke et al. (2023) developed a soft-masking mechanism to selectively update models using domain-specific corpora, which helps maintain general knowledge while enhancing domain performance. Xie et al. (2023) created FinPythia-6.9B, a model adapted through domain-specific pre-training for the financial sector, transforming a general model into a domain expert.

Despite these advancements, existing CPT techniques often require meticulous data curation, and our experiments have demonstrated that even minor duplications in data can lead to significant performance degradation. Other models like InvestLM (Yang et al., 2023) and MedAlpaca (Han et al., 2023) have shown improvements in domain-specific generalization and adaptation with Low Rank Adaptation (LoRA, Hu et al, 2021) fine-tuning but fall short in facilitating extended context reasoning (Anil et al., 2022; Zhou et al., 2023). Our framework, LLM-ADE, diverges from purely enhancing domain-specific accuracy; instead, it focuses on enriching the linguistic and reasoning capabilities of LLMs across varied data inputs.

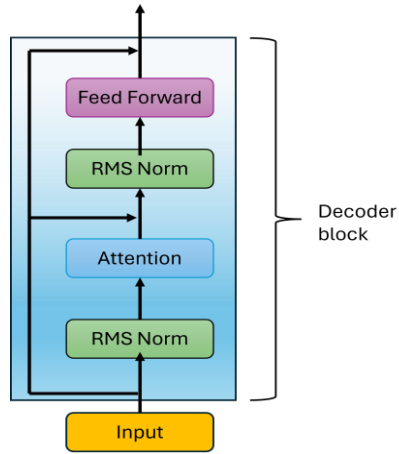

Figure 1: Decoder block

LLM-ADE also incorporates Llama Pro's block expansion (Wu et al., 2024) techniques, which added eight blocks to Llama2-7B and trained on 80 billion tokens for 2,830 GPU hours on Nvidia H800 to create a new foundational model. We focus on optimizing training efficiency on significantly smaller training corpus and fewer resources for broader applicability. Additional innovations include novel block placements and layer modification strategies tailored for each dataset, targeting adaptability and efficiency.

## 3 Methodology

### 3.1 Base Models

The LLM-ADE framework is designed to enrich the capabilities of existing pre-trained large language models (LLMs) by seamlessly integrating new datasets, rather than creating new foundational models. For our experiments, we selected the TinyLlama model developed by Zhang et al. (2024). TinyLlama is an open-source, decoder-only transformer model, that strikes a balance. We specifically chose the 1B parameter TinyLlama due to its balance between computational efficiency and model complexity, ensuring both accessibility and practical applicability in real-world scenarios. TinyLlama's minimal hardware requirements, manageable with a single Nvidia RTX 3090 or L4 GPU, further facilitate this balance.

TinyLlama is a data-saturated model, having been pre-trained on a substantial 3 trillion token corpus. This extensive pre-training meets the rigorous standards set by the Chinchilla optimal thresholds (Hoffman et al., 2022). The high level of data saturation implies that any observed improvements in knowledge retention and processing capabilities within TinyLlama could suggest greater potential advantages for applying the LLM-ADE framework to larger and more complex models. Therefore, by demonstrating performance enhancements in TinyLlama, we aim to highlight the general applicability and effectiveness of the LLM-ADE approach across various model scales.

In terms of architecture, TinyLlama utilizes a decoder-only transformer architecture similar to Llama 2 (Touvon et al., 2023). It incorporates advanced features such as pre-normalization with RMSNorm (Zhang and Sennrich, 2019), SwiGLU activation functions (Shazeer, 2020), and rotary positional embeddings (RoPE, Su et al., 2022), comprising 22 blocks. An illustrative diagram of a decoder-only block is provided in Figure 1.

### 3.2 Block importance

To effectively identify which blocks are critical for adaptation during the continued pre-training phase, we leverage methodologies from recent research in layer-pruning. Specifically, we have adapted an angular distance (AD) metric, based on the work by Gromov et al. (2024) and Men et al. (2024), for assessing the relevance of each block within our model. The angular distance between the inputs of block $i$ and block $i+1$ is calculated using the following formula:

$$AD_i = \frac{X_i^T X_{i+1}}{\|X_i\|_2 \|X_{i+1}\|_2} \quad . \tag{1}$$

This calculation helps identify blocks where significant data processing shifts occur, indicating their importance for adapting to new information. We calculate the inverse cosine of the average angular distance, i.e., $\cos^{-1}(-E[AD_i])$ using an independent 5% of the target dataset. This metric identifies those blocks which undergo the most substantial changes when exposed to new data. Blocks exhibiting the highest average values are then prioritized for modifications, such as selective tuning or expansion, to enhance the model's adaptability and performance on the specific target dataset being integrated.

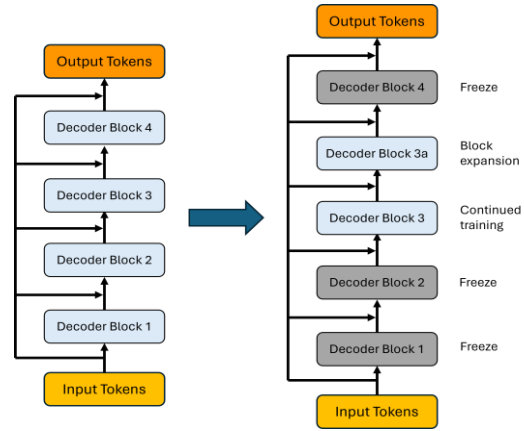

Figure 2: Block modifications

### 3.3 Block Adjustments

To mitigate catastrophic forgetting during architectural modifications, it is essential to ensure that such modifications do not significantly degrade the model's existing knowledge base. To address this, we employ a strategy of freezing all blocks during training except where the angular distance between inputs shows the highest variance. This selective freezing prevents updates to these blocks during backpropagation, thus preserving the weights of most blocks but only updating weights of the blocks where the most data processing is performed. Additionally, we expand the model's capacity by adding new blocks immediately following the critical blocks identified. This is depicted in Figure 2, which illustrates a model architecture after the selection of the third block for further training and expansion.

| Training/Dataset | Avg Improvement | Hellaswag | Winogrande | Piqa | OpenBookQA | bigbench |
|---|---|---|---|---|---|---|
| TinyLlama | | 44.2 | 59.3 | 72.6 | 25.2 | 37.1 |
| CPT 100% Slim Pajama | -12.00 | 31.1 | 50.0 | 55.0 | 12.0 | 30.4 |
| CPT 80% Hermes, 20% SP | -0.02 | 44.9 | 59.3 | 71.5 | 24.8 | 37.8 |
| | | | | | | |
| CPT 90% Hermes, 10% SP | 0.04 | 45.8 | 60.0 | 73.3 | 21.2 | 38.3 |
| CPT 100% Hermes | 0.57 | 45.7 | 59.9 | 72.0 | 25.4 | 38.3 |

Table 1: TinyLlama base and CPT/LoRA training on Hermes and mixed datasets

Initial experiments with block expansion involved a strategy of copying weights from the previous blocks to the new ones (Wu et al., 2024). However, our empirical results indicated slightly improved performance when these newly added blocks were initialized with random weights, which were then scaled to align with the distribution of the existing model weights. This approach eliminates the requirement for high initial learning rates, facilitating a more gradual and effective integration of new information while preserving the integrity of previously acquired knowledge.

## 4 Experiments on General Knowledge Dataset

To evaluate the effectiveness of the LLM-ADE framework, we conducted a series of experiments focusing on continual pre-training and fine-tuning using a general knowledge, general use dataset.

### 4.1 Data

For this study, we used the OpenHermes 2.5 dataset (Teknium, 2023), consisting of 1 million high-quality synthetic samples from instruction and chat data generated by GPT-4. This dataset, covering a broad spectrum of AI-related topics, was distinct from the pre-training data of our base model, TinyLlama. We utilized approximately half of this dataset (500,000 sequences or 200 million tokens) to simulate a realistic dataset size for practical applications, reserving the remainder for testing block importance. Additionally, to evaluate the model's resistance to catastrophic forgetting, we included randomized subsets from the SlimPajama dataset (Soboleva et al., 2023).

### 4.2 Benchmarks

Our model's performance was benchmarked using several well-established general knowledge tests: HellaSwag (Zelles et al., 2019), Winogrande (Sakaguchi et al., 2019), Piqa (Bisk et al., 2019), OpenBookQA((Mihaylov et al., 2018), and BigBench (Srivastava et al., 2022). These evaluations were rerun on our evaluation pipelines using the Language Model Evaluation Harness (Gao et al, 2023), a robust, standardized framework, for consistent comparisons.

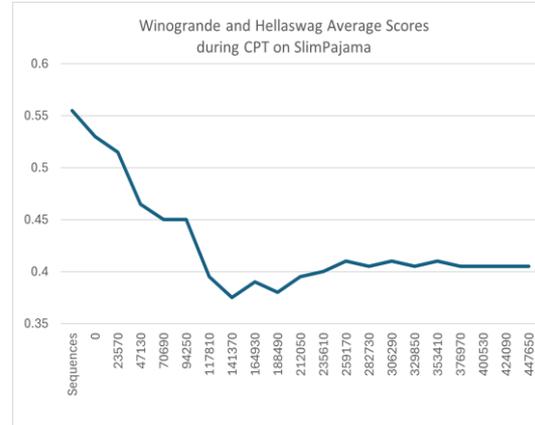

Figure 3: TinyLlama CPT Catastrophic Forgetting with Duplicate Data

### 4.3 CPT and LoRA

We conducted continuous pre-training (CPT) on the OpenHermes 2.5 dataset, testing different learning rates and batch sizes. The optimal settings were found to be a cosine learning rate schedule with a maximum of $4.0 \times 10^{-5}$ and a minimum of $4.0 \times 10^{-6}$, with a batch size of 2 million tokens. LoRA fine-tuning was also applied to the base model using the same learning rates, a batch size of 1 million tokens, r=256, and alpha=512. On a single Nvidia L4 GPU, CPT training time for this dataset required 25 hours.

Our experiments with the SlimPajama dataset allowed us to observe the effects of data duplication: initial CPT on the OpenHermes dataset marginally improved performance by 0.57 points, but incorporating the SlimPajama dataset negated these gains, eliminating most of the gains with only 10% duplicative data and ultimately leading to inferior performance compared to the base model with 20% duplication. While training time was reduced to 15 hours, LoRA fine-tuning did not fare much better, underperforming the scores of the continual training.

Notably, training solely on the SlimPajama dataset led to catastrophic forgetting. Figure 3 illustrates this through a graph showing the performance of TinyLlama on the HellaSwag and Winogrande benchmarks at each 1/20[th] interval of CPT solely on SlimPajama data. The model's performance dropped significantly immediately after the introduction of the SlimPajama data and did not recover throughout the training period.

| Block | 5% | 10% | Full |
|---|---|---|---|
| 1 | 0.87 | 0.87 | 0.87 |
| 2 | 1.67 | 1.68 | 1.68 |
| 3 | 1.44 | 1.44 | 1.44 |
| 4 | 1.36 | 1.36 | 1.36 |
| 5 | 1.35 | 1.35 | 1.35 |
| 6 | 1.16 | 1.16 | 1.16 |
| 7 | 1.50 | 1.50 | 1.50 |
| 8 | 1.61 | 1.61 | 1.61 |
| 9 | 1.36 | 1.36 | 1.36 |
| 10 | 1.34 | 1.34 | 1.34 |
| 11 | 1.38 | 1.38 | 1.38 |
| 12 | 1.51 | 1.51 | 1.51 |
| 13 | 1.44 | 1.44 | 1.44 |
| 14 | 1.09 | 1.09 | 1.09 |
| 15 | 1.26 | 1.26 | 1.26 |
| 16 | 1.01 | 1.01 | 1.01 |
| 17 | 0.86 | 0.86 | 0.86 |
| 18 | 0.84 | 0.84 | 0.84 |
| 19 | 0.87 | 0.87 | 0.87 |
| 20 | 1.22 | 1.22 | 1.22 |
| 21 | 1.42 | 1.42 | 1.42 |

Table 2: Block Importance metrics on different samples of the dataset

### 4.4 LLM-ADE

For block importance testing, we chose to analyze a 5% subset of the target dataset, as the relative rankings of block importance remained consistent across 5%, 10%, and 100% of the dataset (Table 2). The metrics indicated that blocks 2 and 8 were of highest importance. The effectiveness of the LLM-ADE technique, which involves unfreezing and expanding these blocks, is demonstrated in Table 3. While improvements were observed when unfreezing or expanding individual blocks, the most significant enhancements were achieved when both blocks were modified simultaneously, surpassing the results from previous CPT and LoRA configurations of 0.78 points improvement on the base model. The improvements were mostly maintained even with the introduction of duplicative data: even with 20% SlimPajama mix, the LLM-ADE model held 0.50 point impovements. This comparison highlights the benefits of the LLM-ADE approach, particularly when both blocks are unfrozen or expanded, as opposed to solely freezing or expanding.

## 5 Discussion and conclusion

The LLM-ADE framework introduces a novel approach to continual pre-training of large language models, addressing the challenges of efficiently integrating new datasets while mitigating the risks of catastrophic forgetting and double descent. By strategically identifying critical blocks within the model architecture using angular distance metrics, LLM-ADE enables targeted modifications such as selective freezing and block expansion. This approach allows for the effective integration of rapidly updating datsets while preserving the model's existing knowledge base.

| Training/Dataset | Avg Improvement | Hellaswag | Winogrande | Piqa | OpenBookQA | bigbench |
|---|---|---|---|---|---|---|
| LLM-ADE 80% Hermes, 20% SP | 0.50 | 46.0 | 59.3 | 72.4 | 24.4 | 38.8 |
| LLM-ADE 90% Hermes, 10% SP | 0.73 | 46.7 | 60.7 | 73.6 | 21.9 | 39.1 |
| LLM-ADE 100% Hermes | 0.78 | 46.6 | 60.3 | 72.5 | 23.9 | 38.9 |

Table 3: LLM-ADE training on Hermes and mixed datasets

Experiments conducted on the TinyLlama model using the OpenHermes 2.5 dataset illustrate the of improvements of the LLM-ADE technique compared to traditional continuous pre-training (CPT) and LoRA fine-tuning methods. The simultaneous unfreezing and expansion of high-importance blocks yielded the most significant performance improvements, surpassing the results obtained through individual block modifications or alternative training configurations.

Furthermore, LLM-ADE's efficiency in resource utilization marks a step forward in sustainable AI development, aligning with the increasing need for power-efficient and environmentally conscious technology solutions. The successful application of LLM-ADE to the TinyLlama model underlines the framework's potential applicability across various LLMs of different sizes and complexities, suggesting its adaptability to broader use cases in the AI industry.

In summary, LLM-ADE not only meets the rigorous demands of modern AI tasks but also sets a new standard for future developments in the domain of machine learning and AI. It promises to enhance the robustness, flexibility, and efficiency of LLMs, paving the way for more dynamic, adaptable, and efficient models that are capable of evolving in sync with the rapid pace of information change. This framework could potentially revolutionize how we train and maintain state-of-the-art LLMs, making continuous learning a practical and scalable reality